\documentclass[pra,aps,twocolumn,amsmath,amssymb,showpacs]{revtex4}

\usepackage{graphicx}
\usepackage{color}
\usepackage[dvips]{epsfig}

\newcommand{\ppnp}{\phi'(0^+)}
\newcommand{\ppnm}{\phi'(0^-)}
\newcommand{\ppn}{\phi'(0)}
\newcommand{\pnp}{\phi(0^+)}
\newcommand{\pnm}{\phi(0^-)}
\newcommand{\pn}{\phi(0)}
\newcommand{\gb}{g_{\text{B}}}
\newcommand{\gf}{g_{\text{F}}}

\newcommand{\be}{\begin{equation}}
\newcommand{\ee}{\end{equation}}

\begin{document}

\title{Discretized vs. continuous models of p-wave interacting fermions
in 1D}

\author{Dominik Muth and Michael Fleischhauer}
\affiliation{Fachbereich Physik und Forschungszentrum OPTIMAS, Technische Universit\"{a}t
Kaiserslautern, D-67663 Kaiserslautern,
Germany}
\author{Bernd Schmidt}
\affiliation{Institut f\"ur Theoretische Physik,\\
Johann Wolfgang Goethe-Universit\"{a}t Frankfurt,
D-60438 Frankfurt am Main,
Germany}

\begin{abstract}
We present a general mapping between continuous and lattice models of
Bose- and Fermi-gases in one dimension, interacting via local two-body interactions. 
For $s$-wave interacting bosons we arrive at the Bose-Hubbard model in the weakly 
interacting, low density regime. The dual problem of $p$-wave interacting fermions
is mapped to the spin-1/2 XXZ model close to the critical point in the highly polarized regime. 
The mappings are shown to be optimal in the sense that they produce the least error possible for a given 
discretization length. As an application we examine the ground state of a interacting Fermi gas in a harmonic trap, 
calculating numerically real-space and momentum-space distributions as well as two-particle correlations. In the 
analytically known limits the convergence of the results of the lattice model to the continuous one is
shown.
\end{abstract}


\maketitle

\section{introduction}

Triggered by the recent successes in the experimental realization 
of strongly interacting atomic quantum gases
in one spatial dimensional (1D) \cite{Kinoshita2004,Paredes2004,Gunter2005,Hofferberth2007,Haller2009}
there is an increasing interest in
the theoretical description of these systems beyond the mean field level.
Model hamiltonians describing homogeneous 1D quantum gases with contact 
interaction are often integrable by means of Bethe Ansatz \cite{Bethe1931,Lieb1963a,Korepin1993,Gaudin1983}. 
In practice, however,
only a small number of quantities can actually be obtained from Bethe Ansatz 
or explicit calculations are restricted to a small number of particles. 
Properties associated with low energy or long wavelength
excitations can, to very good approximation, be described by bosonization techniques \cite{Giamarchi2003}.
For more general problems one has to rely on numerical
techniques such as the density matrix renormalization group (DMRG) \cite{White1992,Schollwoeck2005} or
the related time evolving block decimation (TEBD) \cite{Vidal2003,Vidal2004}. Both have originally
been developed for lattice models and thus in order to apply them to continuous
systems requires a proper mapping between the true continuum model and a lattice
approximation. In fact any numerical technique describing a continuos
system relies on some sort of discretization.
Here we consider massive bosonic or fermionic particles with contact
interactions. Only two types of contact interaction potentials are allowed for
identical, nonrelativistic particles, representing either bosons with s-wave interactions
or fermions with p-wave interactions. Both systems are dual and can be mapped onto
each other by the well-known boson-fermion mapping \cite{Girardeau1960,Cheon1999}. 
A proper discretization of 1D bosons with s-wave interaction is
straight forward and has been used quite successfully to calculate
ground-state \cite{Schmidt2007}, finite temperature \cite{Schmidt2005}, as well as dynamical problems \cite{Muth2009}
for trapped 1D   gases. For p-wave interacting fermions a similar, straight forward discretization 
fails however, as can be seen when comparing numerical results using such a model 
with those obtained from the bosonic Hamiltonian after the boson-fermion
mapping. Using a general approach to quantum gases in 1D with contact interaction \cite{Schmidt2009} we 
here derive a proper mapping between continuous model and lattice approximation. We show in particular
that p-wave interacting fermions are mapped to the critical spin 1/2 XXZ model. By virtue
of the boson-fermion mapping the same can be done for s-wave interacting bosons, thus
maintaining integrability in the map between continuous and discretized models. As an
application we calculate the real-space and momentum-space densities of the ground
state of a p-wave interacting Fermi gas in a harmonic trap, as well as local
and non-local two particle correlations in real space. To prove the validity of the discretized
fermion model we compare the numerical results with those obtained from the
dual bosonic model as well as with Bethe ansatz solutions when available.

\section{1D quantum gases with general contact interactions}

We here consider quantum gases, that are fully described by their two particle Hamiltonian, i.e.,
the Hamiltonian is a sum of the form
\begin{equation}
 H = -\frac{1}{2}\sum_j\partial_{x_j}^2 +\sum_{i<j}V(x_i-x_j).
\end{equation}
Additionally we require that the true interaction potential can be approximated 
by a local pseudo-potential, i.e. it vanishes for $x_i\ne x_j$. Since we are in one dimension, this 
leads to the exact integrability of these models in the case of translational invariance \cite{Lieb1963a} using coordinate Bethe ansatz \cite{Bethe1931,Gaudin1983}. 

For deriving a discretized Hamiltonian, it is sufficient to consider the relative wave 
function $\phi(x=x_1-x_2)$ of just {\it two} particles. The Hamiltonian then reads
\begin{equation}
 H = -\partial_{x}^2 + V(x)
\end{equation}
where we have dropped the term corresponding to the freely evolving center of mass.

The continuous two-particle case has been analyzed by Cheon and Shigehara \cite{Cheon1998,Cheon1999}. 
The local pseudo-potential $V$ is fully described by a boundary condition on $\phi$ at $x=0$: Since $\phi$ fulfills the free Schr\"odinger equation away from $0$, it must have a discontinuity at the origin as an effect of the interaction. Thus we see that
\begin{equation}\label{eq:kinetic}
 \partial_{x}^2\phi(x) = \left\{\begin{array}{cl}
                                \phi''(x) & x\ne 0\\
                                \delta(x)\left[\ppnp-\ppnm\right]+&\\\quad +\ \delta'(x)\left[\pnp-\pnm\right] & x=0.
                               \end{array}
\right.
\end{equation}
In the case of distinguishable or spinful \cite{Girardeau2004} particles both singular terms contribute. Due to symmetry, the term proportional to the delta function $\delta$ can only be nonzero for bosons, 
while the $\delta'$ term exists only for fermions. I.e. we have for bosons
\begin{equation}\label{eq:kinetic-b}
  \partial_{x}^2\phi(x) = \left\{\begin{array}{cl}
                                \phi''(x) & x\ne 0\\
                                2\delta(x)\phi^\prime(0) & x=0.
                               \end{array}
\right.
\end{equation}
and for fermions
\begin{equation}\label{eq:kinetic-f}
 \partial_{x}^2\phi(x) = \left\{\begin{array}{cl}
                                \phi''(x) & x\ne 0\\
                                \ 2\delta'(x)\phi(0) & x=0.
                               \end{array}
\right.
\end{equation}
In order to get proper eigenstates (i.e. without any singular contribution), the
pseudo-potential $V$ acting on the wave-function must absorb the singular contributions from the kinetic energy. 
Thus the only possible form of a local pseudo-potential for bosons is
$V_B\phi = \gb\delta(x)\phi(0)$, while that for fermions reads $V_F\phi = -\gf\delta'(x)\phi'(0)$. Note that $\phi$ ($\phi'$) is continuous at 0 for bosons (fermions). These two possibilities represent the well known cases, where the particle interact either by s-wave scattering only or by p-wave scattering only, and the interaction strength corresponds to the scattering length, respectively scattering volume, which are the only free parameters left.

Since all wave functions must have the respective symmetry, we can restrict ourselves in the following to the $x>0$ sector. We will write $\pn$ for $\lim_{x\rightarrow0^+}\phi(x)$ and $\ppn$ for $\lim_{x\rightarrow0^+}\phi'(x)$. The above shows that $V$ imposes a boundary condition on every proper wave function: 
\begin{equation}\label{eq:bc}
\begin{array}{ll}
\ppn = \frac{\gb}{2}\pn & \quad\textrm{bosons},\cr
\ppn = -\frac{2}{\gf}\pn & \quad\textrm{fermions}.
\end{array}
\end{equation}
Eqs.(\ref{eq:kinetic}) and (\ref{eq:bc}) reveal a one-to-one mapping between the two cases, i.e., every solution for the bosonic problem yields a solution for the fermionic problem with $\gb = -4/\gf$ by symmetrizing the wave function and vice versa.

At this point we emphasize, that boundary conditions of the above form are the only ones that are equivalent to a local potential \cite{Seba1986a,Cheon1998}. While boundary conditions involving higher order derivatives can be taken into account to describe experimental realizations using cold gases in quasi 1D traps \cite{Imambekov2009}, the necessarily require finite range potentials and cannot be described fully by local pseudo-potentials.

\section{discretization}

The treatment of continuous gases in one-dimension using numerical techniques requires a proper discretization. That is we approximate the two-particle wave function $\phi(x) \in L^2(\mathbb{R})$ by a complex number $\phi_j \in \ell^2(\mathbb{Z})$, where the integer index $j$ describes the discretized relative coordinate $x=x_1-x_2$.
We interpret $\vert\phi_j^2\vert$ as the probability to find the two particles between $(j-\frac12)\Delta x$ and $(j+\frac12)\Delta x$. In order to apply numerical methods such as DMRG or TEBD
\cite{Vidal2004,Vidal2003} efficiently, it is favorable to have local or at most nearest neighbor interactions
in the lattice approximation of the continuous model.
It will turn out, that the above systems can all be discretized using such nearest neighbor interactions only.

We start with the kinetic term, that can be approximated by
\begin{equation}
 \partial_x^2 \mapsto \frac{\phi_{j-1}-2\phi_j+\phi_{j+1}}{\Delta x^2}.
\end{equation}
In what follows, we will derive two distinct discretizations: first for the bosons, where we allow for double occupied lattice sites and can therefore use on-site interactions to reproduce the boundary conditions (\ref{eq:bc}), and then for fermions, where double occupation is forbidden by the Pauli principle and interactions between neighbors are necessary
in the lattice model. Note however, that both descriptions are equivalent due to the Bose Fermi mapping in the continuum
limit.

\subsection{bosonic mapping}
%
\noindent 
In the lattice approximation the kinetic-energy term, Eq.(\ref{eq:kinetic}) reads
\begin{equation}
 \partial_{x}^2\phi(x) = \left\{\begin{array}{cl}
\frac{\phi_{j-1}-2\phi_j+\phi_{j+1}}{\Delta x^2}&j>0 \\
\frac{2(\phi_1 -\phi_0)}{\Delta x^2} & j=0
                               \end{array}
\right.
\end{equation}
Thus assuming a local contact interaction only, we find for the bosons
\begin{equation}
 (H\phi)_j = \left\{\begin{array}{cl}
-\frac{\phi_{j-1}-2\phi_j+\phi_{j+1}}{\Delta x^2}&j>0 \\
U\phi_0-\frac{2\phi_1-2\phi_0}{\Delta x^2}&j=0
                    \end{array}\right..
\end{equation}
In order to determine the value of $U$, we assume, that it can be expressed as a series in $\Delta x$ and evaluate the stationary Schr\"odinger equation $(H\phi)_j-E\phi_j=0$ at $j=0$. Reexpressing $\phi_1$ in terms of $\phi(0)$ by means of
the discretized version of the contact condition (\ref{eq:bc})
\begin{equation}
 \phi_1 = \pn + \Delta x\underbrace{\ppn}_{=\frac{\gb}2\pn} +\frac{\Delta x^2}2\underbrace{\phi''(0)}_{=-E\pn}+\dots,
\end{equation}
we arrive at
\begin{eqnarray}
 0 &=& (H\phi)_{j=0}-E\phi_{j=0} \nonumber\\
&=& U\pn +\frac{2\pn}{\Delta x^2} - E\pn - \frac2{\Delta x^2} \times \\&\times&\left(\pn+\Delta x\frac{\gb}2\pn-\frac{1}{2}\Delta x^2E\phi(0)+\mathcal{O}(\Delta x^3)\right)\nonumber  .
\end{eqnarray}
Equating orders gives
\begin{equation}\label{eq:uresult}
 U = \frac{\gb}{\Delta x} + \mathcal{O}(\Delta x).
\end{equation}
The constant term vanishes, since $-\partial_x^2\phi=E\phi$ for any eigenstate. The higher orders $\mathcal{O}(\Delta x)$ contain $E$ and would thus not be independent on the eigenvalue. This is perfectly consistent, since discretizations will only work a long as the lattice spacing is much smaller than all relevant (wave) lengths in the system. Thus the lowest order in (\ref{eq:uresult}) is already optimal. There are no higher order corrections possible for a general state.

We can now easily write down the corresponding {\it many} particle Hamiltonian for the case of indistinguishable bosons in {\it absolute} coordinates, represented by an integer index $i$  and in second quantization:
\begin{equation}
 H = \sum_i\left[-J(a_i^\dagger a_{i+1}+h.a.) +\frac{U}{2}a_i^\dagger a_i^\dagger a_{i} a_{i} + V_i 
a^\dagger_{i}a_i\right].
\end{equation}
Here $a_i$ is the bosonic annihilator at site $i$ and $V_i$ introduces an additional external potential in the obvious way. So not surprisingly we have arrived at the Bose-Hubbard Hamiltonian as a lattice approximation
to 1D bosons with s-wave interaction. Since $\Delta x$ must be smaller than all relevant length scales, we are
however in the low-filling and weak-interaction limits $U\ll J=\frac{1}{2\Delta x^2}$ \footnote{In the case of ground state calculations as done in section \ref{sec:fgas} we actually achieve good results even before $J$ exceeds $U$. However for non equilibrium dynamics \cite{Muth2009} it can become crucial that the bandwidth proportional to $J$ is large compared to the pairing energy $U$.}. This does of course not imply that the corresponding Lieb-Liniger gas is in the weakly interacting regime. This result might seem trivial, since we can also directly get it by substituting the field operator in the continuous model: $\Psi(j\Delta x)\mapsto\frac{a_j}{\sqrt{\Delta x}}$ \cite{Schmidt2007}. However, this simple 
and naive discretization does not work in the fermionic case we are going to discuss now.

\subsection{fermionic mapping}

\noindent For fermions the kinetic-energy term, Eq.(\ref{eq:kinetic}) reads
in lattice approximation
\begin{equation}
 \partial_{x}^2\phi(x) = \left\{\begin{array}{cl}
\frac{\phi_{j-1}-2\phi_j+\phi_{j+1}}{\Delta x^2}&j>1 \\
\frac{\phi_2-2\phi_1}{\Delta x^2}&j=1\\
0 & j=0
                               \end{array}
\right.
\end{equation}
Due to the anti-symmetry of the wave-function
$\phi_0$ must vanish, i.e. the simplest way interactions come into the
lattice model is for nearest neighbors. Thus we 
write for the Hamiltonian 
\begin{equation}
 (H\phi)_j = \left\{\begin{array}{cl}-\frac{\phi_{j-1}-2\phi_j+\phi_{j+1}}{\Delta x^2}&j>1
                    \\
                     B\phi_1-\frac{\phi_2-2\phi_1}{\Delta x^2}&j=1\\
 0&j=0                    \end{array}\right.
\end{equation}
To obtain the value of $B$ we proceed as in the case of bosons.
As will be seen later on it is most convenient to expand $B$ in a series in the following way:
\begin{equation}\label{eq:bexpansion}
 \frac{1}{B} = \Delta x^2\left(B^{(2)}+\Delta xB^{(3)}+\mathcal{O}(\Delta x^2)\right).
\end{equation} 
Now the stationary Schr\"odinger equation for $j=1$ yields
\begin{eqnarray}
&&0=1-\frac2{\gf}\Delta x - \frac{\Delta x^2}2E+\mathcal{O}(\Delta x^3) +\\
&&+\left(B^{(2)}+\Delta xB^{(3)}+{\Delta x^2}B^{(4)}+ \mathcal{O}(\Delta x^2)\right)\left[1 + \mathcal{O}(\Delta x^3)\right].\nonumber
\end{eqnarray}
Equating orders results in
\begin{equation}\label{eq:bresult}
 B^{(2)} = -1,\quad B^{(3)} = \frac{2}{\gf},\quad B^{(4)} = \frac12E.
\end{equation}
Note that his time the interaction appears only in the \emph{second} lowest order, which can not be described by a simple substitution formula. The next higher order contained in $\mathcal{O}(\Delta x^2)$ does not vanish, but depends again on the energy as expected. If we had chosen a straightforward expansion of $B$ instead of (\ref{eq:bexpansion}), the next order after the one that introduces the interaction would have contained again the interaction parameter:
\begin{equation}
 B = -\frac{1}{\Delta x^2} -\frac{2}{g_F\Delta x} -\frac{4}{g_F^2}+\frac{E}{2} +\mathcal{O}(\Delta x).
\end{equation}
Neglecting this term would therefore introduce a larger error than in the chosen expansion (\ref{eq:bexpansion}). In fact the low energy scattering properties would be reproduced only to one order less. For the bosons this problem did not occur (\ref{eq:uresult}). From (\ref{eq:bexpansion}) we read that the optimal result in the fermionic case is
\begin{equation}\label{eq:B}
 B = -\frac{1}{\Delta x^2}\left(\frac{1}{1-\frac{2\Delta x}{\gf}
}\right).
\end{equation}
The corresponding many-body Hamiltonian for indistinguishable fermions reads
\begin{equation}\label{eq:hfermilattice}
 H = \sum_i\left[-J(c_i^\dagger c_{i+1}+h.a.) +Bc_i^\dagger c_{i}c_{i+1}^\dagger c_{i+1} + V_i c_i^\dagger 
c^\dagger_{i}\right],
\end{equation}
where now $c_i$ is a fermionic annihilator at site $i$. Eq. (\ref{eq:hfermilattice}) describes spin polarized lattice
fermions with hopping $J$ and nearest-neighbor interaction $B$. 
In contrast to the bosonic case, Eq.(\ref{eq:bresult}), where the correct discretized model could be obtained from
the continuum Hamiltonian just by setting $\Psi(x) \rightarrow a_i /\sqrt{\Delta x}$, we now see from (\ref{eq:hfermilattice}) and (\ref{eq:B}) that a similar naive and straight-forward discretization fails in the
case of $p$-wave interacting fermions. 

The failure of a naive discretization of the fermionic Hamiltonian becomes transparent 
if we map this model to that of a spin lattice: 
Using the Jordan-Wigner transformation
\begin{equation}
\sigma_i^+ = \exp\Bigl\{i\pi\sum_{l<i}  c_l^\dagger  c_l\Bigr\}\,  c_i,\qquad  \sigma_i^z =
2 c_i^\dagger c_i -1
\end{equation}
(\ref{eq:hfermilattice}) can be mapped to the spin-1/2
XXZ model in an external magnetic field
\begin{eqnarray}
 H &=& \sum_i\Big\{-\frac1{4\Delta x^2}\big(\sigma_i^x\sigma_{i+1}^x+\sigma_i^y\sigma_{i+1}^y+\nonumber\\
&+&\Delta(\sigma_i^z+1)(\sigma_{i+1}^z+1)\big)+V_i\sigma_i^z\Big\},
\end{eqnarray}
where the anisotropy parameter defining the XXZ model is $\Delta=-1/[1-\frac{2\Delta x}{\gf}
]$.

There is an easy way to see that these mappings are quite physical by considering the ground states: The repulsive Bose gas ($\gb>0$) maps to the repulsive ($U>0$) Bose-Hubbard model in the super fluid, low filling regime, which has an obviously gas like ground state. The same is true for the corresponding attractively interacting ($\gf<0$) Fermi gas, which maps to the ferromagnetic XXZ model which, due to the specific form of the interaction parameter in the
discretized fermion model, Eq.(\ref{eq:B}), is always in the critical regime close to the transition point ($\Delta^{\underrightarrow{\Delta x\rightarrow0}}-1^+)$. A naive discretization would have lead to an anisotropy
parameter that could cross the border to the gapped phase, which is clearly unphysical.

In the attractive Bose gas, bound states emerge, that lead to a collapse of the ground state as it is of course also true in the Bose Hubbard model for $U<0$. On the fermionic side, this collapse can be also observed, as for $\Delta<-1$ the XXZ model has a ferromagnetically ordered ground state, which leads to phase separation in the case of fixed magnetization.

Note that we call the Fermi gas repulsively interacting if $\gf>0$, although $B$ is negative in this case as well, and although there exist bound states, who's binding energy actually diverges as $\gf\rightarrow0^+$, as is immediately clear from the Bose Fermi mapping in the continuous case.

\begin{figure}[htb]
	\centering
	\epsfig{file=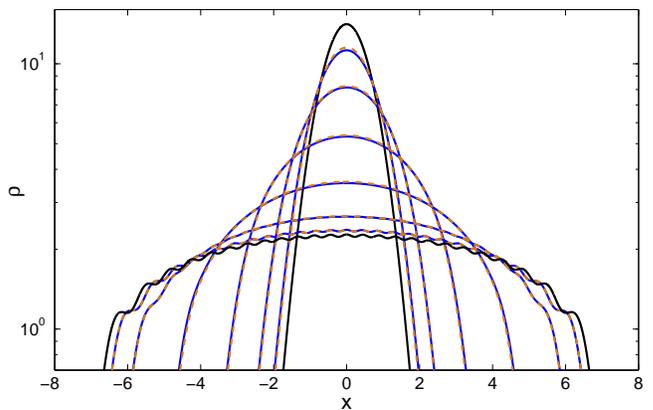, width=\columnwidth}
      \caption{(Color online) Local density distribution of the interacting Fermi or Bose gas. The (orange) dashed lines show 
results obtained by Bose-Fermi mapping and solving the Bose Hubbard lattice model, the (blue) continuous lines correspond to the XXZ discretization. The interaction strength $\gf$ is $-51.2, -12.8, -3.2, -0.8, -0.2, -0.05$ from the narrow to the broad distributions. The solid black lines show the limiting cases of free fermions (broad) and infinitely strong interacting fermions (narrow, corresponds to free bosons). The calculations are done for $\Delta x=\frac{1}{64}$. One
recognizes perfect agreement between the fermionic and bosonic discretization approaches. Note that both Fermions and Bosons with corresponding interaction show the same local density, since the quantity is invariant under the Bose-Fermi-mapping.}
      \label{fig:ld}
\end{figure}

\section{the interacting Fermi gas in a harmonic trap}\label{sec:fgas}

\begin{figure}[htb]
	\centering
	\epsfig{file=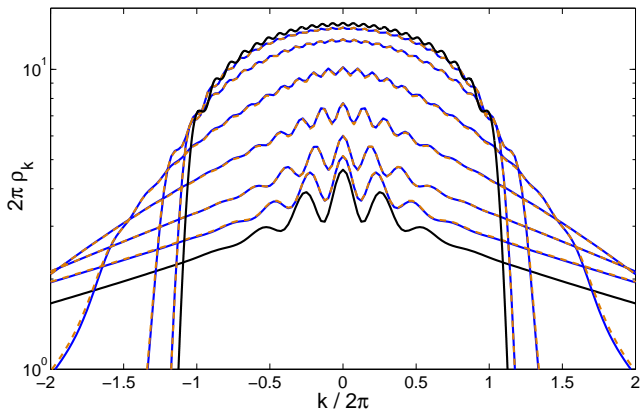, width=\columnwidth}
      \caption{(Color online) Momentum distribution of the interacting Fermi gas. Dashed (orange) lines show results via
 the Bose Hubbard discretization, solid (blue) lines correspond to XXZ discretization. The interaction strength $\gf$ is $-51.2, -12.8, -3.2, -0.8, -0.2, -0.05$ from the broad to the narrow distributions. Solid (black) lines show the limiting cases of free fermions (narrow) and infinitely strong interacting fermions (broad, calculated from the formula given in \cite{Bender2005}). The calculations are done for $\Delta x=\frac{1}{64}$.
Again there is perfect agreement between bosonic and fermionic discretization.}
      \label{fig:md}
\end{figure}

We now apply our method to the interacting Fermi gas in a harmonic trap,
%
\begin{eqnarray} H &=& -\frac{1}{2}\sum_{i=1}^N \partial_{x_i}^2\nonumber\\
 &-&\frac{\gf}{2}\sum_{j<i}\delta'(x_j-x_i)\left.\left(\partial_{x_j}-\partial_{x_i}\right)\right\vert_{x_j=x_{i^{+}}}\nonumber\\
 &+&\sum_{i=1}^N\frac12x_i^2.
\end{eqnarray}
We here chose the trap length to set the length scale. For $\gf=-\infty$ the system is called a fermionic Tonks-Girardeau gas \cite{Girardeau1960,Girardeau2006,Minguzzi2006}. It can be treated analytically, since it maps to free bosons under the Bose Fermi mapping. E.g., the momentum distribution is known for arbitrary particle numbers \cite{Bender2005}. It is of special experimental relevance, since it is equivalent to the density distribution measured in a time-of-flight experiment. However for intermediate interaction strength numerical calculations are required, which we are now able to do.

First we note, that we now have two options to discretized the model. Direct discretization will yield the XXZ Hamiltonian, while a Bose Fermi mapping will result in the Bose Hubbard Hamiltonian. Both methods of 
course have to produce exactly the same results.

Fig. \ref{fig:ld} shows the spatial density distribution in the ground state for $N=25$ particles, i.e.,
\begin{equation}
 \rho(x) = \int dx_2\dots dx_N \left\vert\phi(x, x_2, \dots, x_N)\right\vert,
\end{equation}
which is approximated by the discretized system as the diagonal elements of $\langle a_i^\dagger a_j \rangle$. The ground state of the discretized system is calculated using a TEBD code and an imaginary time evolution, which has already been applied successfully to calculate the phase diagram of a disordered Bose Hubbard model \cite{Muth2008}. The interaction strength is varied all the way from the free fermion regime to the regime of the fermionic Tonks-Girardeau gas. The density distribution changes accordingly from the profile of the free fermions, showing characteristic Friedel oscillations, to a narrow Gaussian peak for the fermionic Tonks-Girardeau gas. Note that the Bose Fermi mapping does not affect the local density, so the curves are the same for the corresponding bosonic system. I.e. the density distribution 
in the fermionic Tonks-Girardeau regime is identical to that of a condensate of non-interacting bosons.
The curves obtained from the bosonic and fermionic lattice models are virtually indistinguishable which shows that
both approaches are consistent.

\begin{figure}[htb]
	\centering
	\epsfig{file=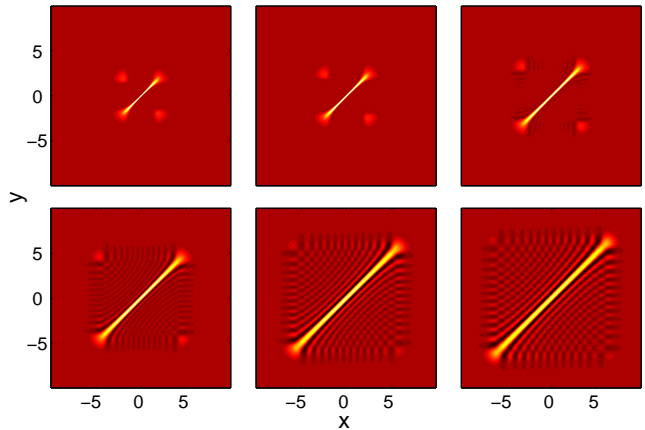, width=\columnwidth}
      \caption{(Color online) single particle density matrix of the interacting Fermi gas calculated using XXZ discretization. Light regions are positive, dark regions negative. The interaction strength $\gf$ is $-51.2, -12.8$, and $-3.2$ (upper row) and $-0.8, -0.2$, and $-0.05$ (lower row). Remember that the cloud size is independent of the particle number towards the fermionic Tonks limit (because there is condensation in the bosonic picture) while it grow as $\sqrt{N}$ for free fermions.}
      \label{fig:densmat}
\end{figure}

The corresponding momentum distribution for the fermions,
\begin{equation}
 \rho_k(k) = \int dk_2\dots dk_N \left\vert\phi(k, k_2, \dots, k_N)\right\vert,
\end{equation}
which is quite different from that of the bosons, is shown in Fig. \ref{fig:md}. It was obtained from the discretized wave function as the diagonal elements of the Fourier transform of $\langle a_i^\dagger a_j \rangle$. Again perfect
agreement between the bosonic and fermionic lattice approximations can be seen.
In accordance with physical intuition invoking the uncertainty relation and Pauli principle, the momentum distribution broadens as the real space distribution narrows. While for the free particles, real and momentum space description coincide for the harmonic oscillator the Friedel oscillations are deformed gradually towards the result for the fermionic Tonks-Girardeau gas calculated e.g. by Bender et al. \cite{Bender2005}. The oscillations that remain in this limit are effects from the finite number of particles. They vanish as $1/N$ as can be seen from a Taylor expansion in $1/N$ of the 
expressions given in \cite{Bender2005} for the Fermi-Tonks case.

\begin{figure*}[htb]
a)\epsfig{file=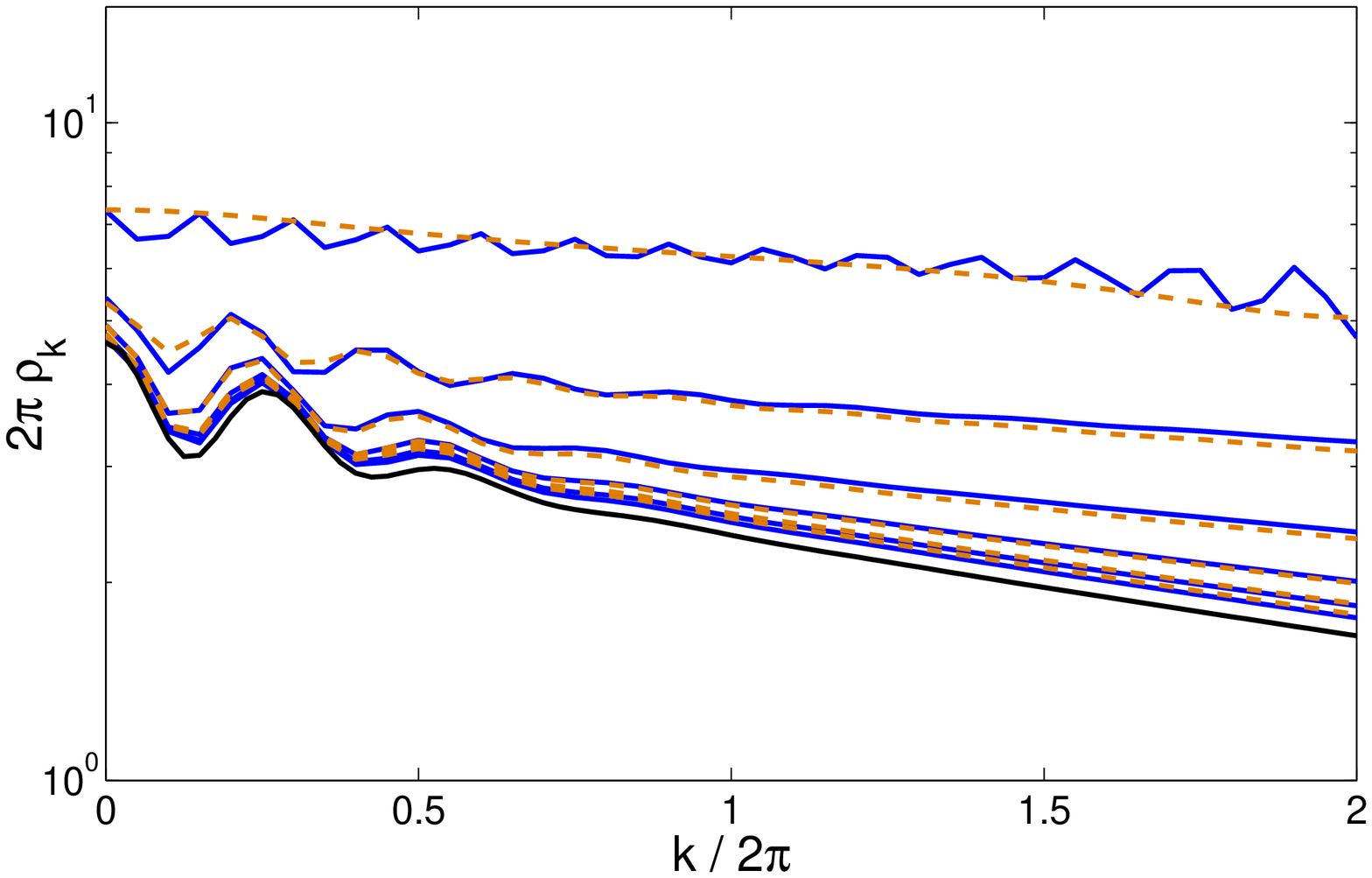, width=.65\columnwidth}
b)\epsfig{file=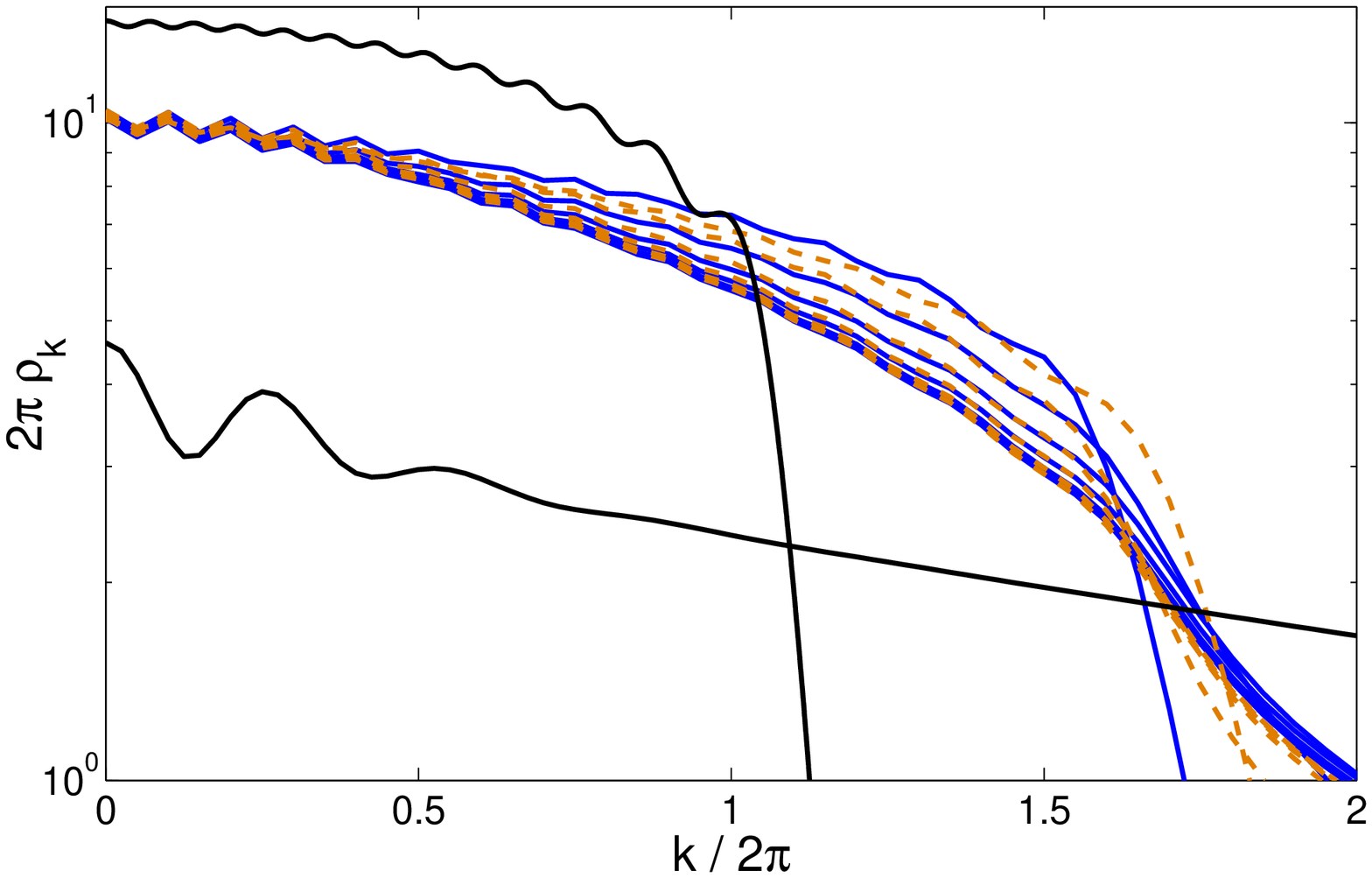, width=.65\columnwidth}
c)\epsfig{file=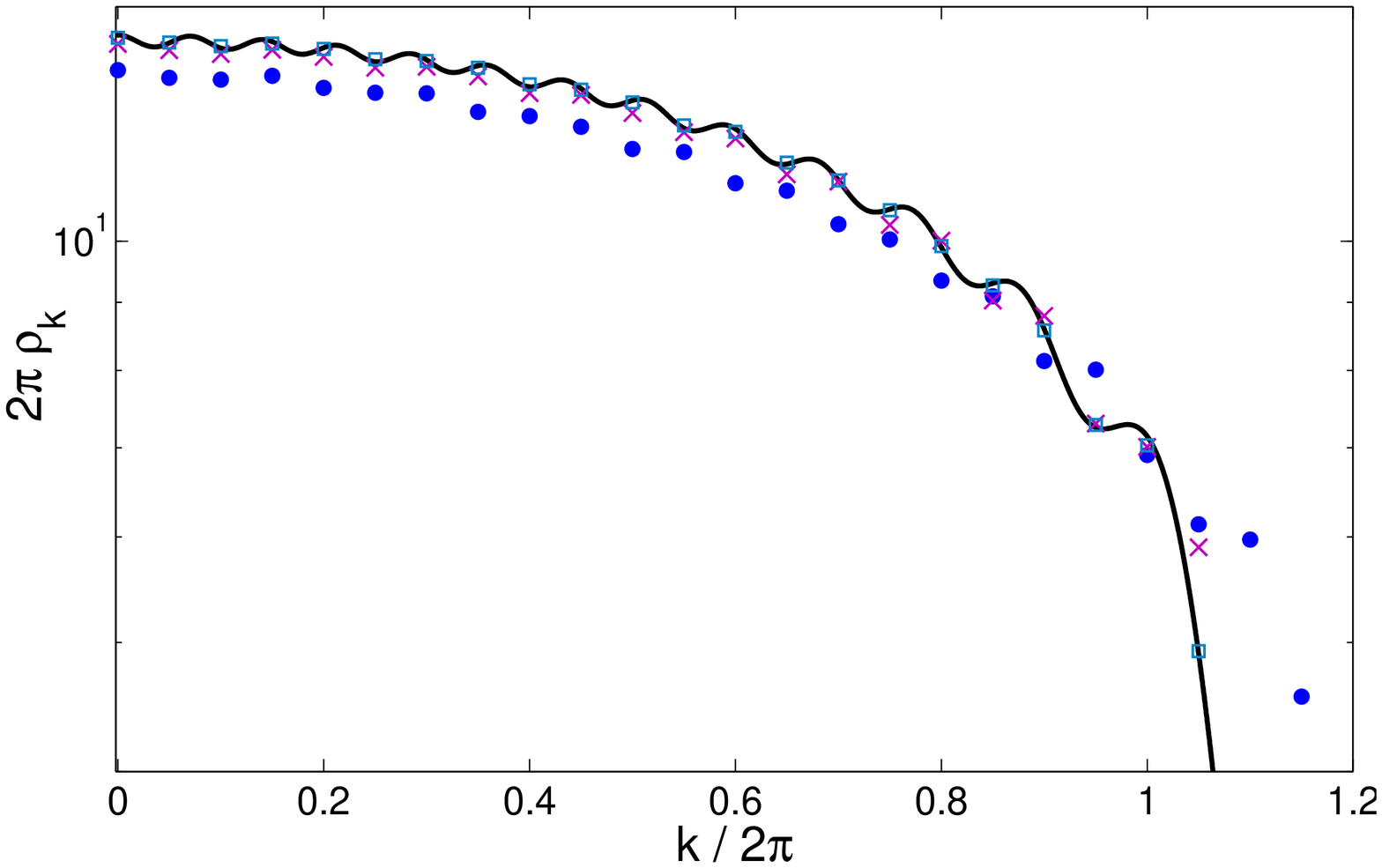, width=.65\columnwidth}
      \caption{(Color online) momentum space distribution of the Fermi gas showing convergence of the method with discretization for a) the Fermi Tonks limit, b) $\gf = -0.8$, and c) the free fermionic case. Again in a) and b) dashed (orange) lines show results via
 the Bose Hubbard discretization, solid (blue) lines correspond to XXZ discretization. a) Results are shown for $\Delta x = \frac{1}{4}, \frac{1}{8}, \frac{1}{16}, \frac{1}{32}, \frac{1}{64}, \frac{1}{128}$. As the grid gets finer, both discretization formulas converge to the exact result (black line). b) The same discretizations are used as in a) and we again observe convergence of both formulas towards a common limit, which is in this case not known analytically. The black lines are those showing up in a) and c) respectively and are for orientation. c) Note that in this case there is no sense in distinguishing the two formulas, since implementing $U=\infty$ always means excluding double occupation of sites by bosons which is immediately equivalent to simulating free fermions. We here only $\Delta x = \frac{1}{4}$ (circles), $\frac{1}{8}$ (crosses), $\frac{1}{128}$ (squares) to avoid confusion since the lines converge quite quickly. Although the squares sit perfectly on top of the exact result (black lines) they are not spaced densely enough to resolve the Friedel oscillations. This would require a lattice that extends across a region in space much larger than $N$ oscillator length where we have chosen to restrict the calculation to $20$ oscillator length to speed it up.}
      \label{fig:conv}
\end{figure*}

In Fig. \ref{fig:densmat} we have plotted the complete
single particle density matrix
\begin{equation}
 \rho(x,y) = \int dx_2\dots dx_N \phi^*(x, x_2, \dots)\phi(y, x_2, \dots)
\end{equation}
for different interaction strength, starting from the
Fermi-Tonks limit to the case of free fermions. One clearly recognizes two small off-diagonal peaks
for larger interaction strength. The weight of these peaks, which are responsible for the
oscillations in the momentum distribution, Fig. \ref{fig:md}, to the remaining part near the diagonal is
$\frac{1}{N}$, as can bee seen from analyzing the limiting case numerically, which can be done for much larger $N$ also. The sign of the peaks is positive only if $N$ is odd and negative for even $N$, so the momentum distributions in Fig. \ref{fig:md} would show a minimum at $k=0$ for all interaction strength if $N$ was chosen even instead of $25$.

\begin{figure}[htb]
	\centering
	\epsfig{file=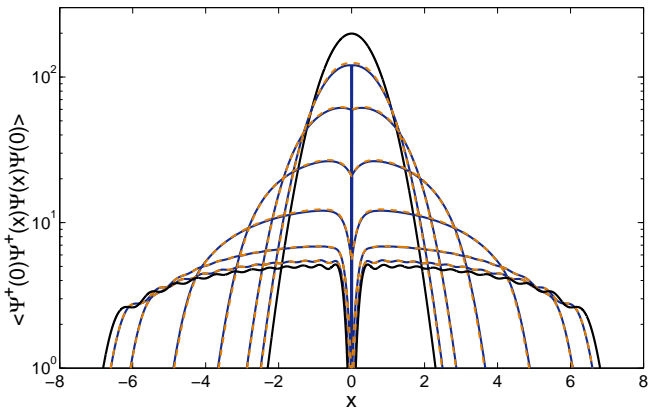, width=\columnwidth}
      \caption{(Color online) density-density correlations of the interacting Fermi or Bose gas. The (orange) dashed lines show 
results obtained by Bose-Fermi mapping and solving the Bose Hubbard lattice model, the (blue) continuous lines correspond to the XXZ discretization. The interaction strength $\gf$ is $-51.2, -12.8, -3.2, -0.8, -0.2, -0.05$ from the narrow to the broad distributions. The solid black lines show the limiting cases of free fermions (broad) and infinitely strong interacting fermions (narrow, corresponds to free bosons). The calculations are done for $\Delta x=\frac{1}{64}$. One
recognizes perfect agreement between the fermionic and bosonic discretization approaches apart from $x=0$ (see text). Note that both Fermions and Bosons with corresponding interaction show the same density-density correlations, since the quantity is invariant under the Bose-Fermi-mapping.}
      \label{fig:g2}
\end{figure}

On first glance it may seem surprising that a mapping of a continuous, Bethe-Ansatz integrable Hamiltonian
such as the Lieb-Liniger model to the non-integrable Bose-Hubbard model should produce accurate
results. However, since the Lieb Liniger gas is dual to $p$-wave interacting fermions, as shown here
its lattice approximation is equivalent to the spin 1/2 XXZ model, which is again Bethe-Ansatz integrable.
Furthermore full recovery of the properties of the continuous model can of course only be expected in the
limit $\Delta x \to 0$. In Fig.\ref{fig:conv} we have shown the momentum distribution of $p$-wave interacting
fermions for decreasing discretization length $\Delta x$ for three different values of the interaction
strength. One clearly recognizes convergence of the results as $\Delta x\to 0$. In the two analytically
tractable cases of a free fermion gas and an the Fermi-Tonks gas the curves approach quickly the
exact ones. 

As a final application we calculate the real-space two-particle correlations in
a trap. The corresponding results are shown in Fig. \ref{fig:g2}. Again the (blue)
solid lines are obtained from the fermionic lattice model and the dashed (orange)
lines from the dual bosonic model. Due to Pauli exclusion $g^{(2)}(0)=0$ and
there is a pronounced dip in the $g^{(2)}$ near the origin for non interacting 
or weakly attractive fermions, while we see again Friedel oscillations for larger inter particle distances. In the dual bosonic case the dip is enforced by a strong
repulsive interaction. As the fermionic attraction is increased, the depth of this
dip is decreased. There is a smooth transition to the perfect Gaussian shape expected for the free bosons in the case of strongly interacting fermions.

Outside the point where the particle positions coincide both discretization formulas give the same result. There is a discontinuity maintaining $g^{(2)}(0)=0$ for the fermions, enforced by the
symmetry of the wave functions. It should be noted that this singular jump is not
reproduced in the dual bosonic model. This is because the duality mapping
of the discretized models is only valid for two particles at {\it different} lattice
sites and the dual bosonic model can only be used to calculate multi-particle correlations
of fermions at pairwise different locations.

Finally we note that using the discretization formulas (\ref{eq:uresult}) and (\ref{eq:bresult}) one can of course also calculate other many body properties like off diagonal order \cite{Minguzzi2006} using TEBD for larger systems. The method was also used to calculate out-of equilibrium dynamics for bosonic gases in the repulsive \cite{Muth2009} as well as attractive regime \cite{MuthFleischhauer_inprep}.

Special thanks go to Anna Minguzzi for stimulating discussions that have lead to this work. The authors would also
like to thank Maxim Olshanii and Fabian Grusdt for valuable input. Finally the financial support of the graduate
school of excellence MAINZ/MATCOR and the Sonderforschungsbereich TR49 are gratefully acknowledged.


\end{document}